\documentclass[a4paper, 12pt]{article}

\usepackage{amsmath,amssymb,amsxtra,epsfig,subfigure}
\usepackage{graphics}
\renewcommand{\vec}[1]{\boldsymbol{#1}}

\numberwithin{equation}{section}

\begin{document}

\title{The stress field in a pulled cork and\\
some subtle points in the \\
semi-inverse method of nonlinear elasticity}
\author{Riccardo De Pascalis, Michel Destrade, Giuseppe Saccomandi}
\date{2006}
\maketitle

\begin{abstract}

In an attempt to describe cork-pulling, we model a cork as an incompressible
rubber-like material and consider that it is subject to a helical shear
deformation superimposed onto a shrink fit and a simple torsion. It turns
out that this deformation field provides an insight into the possible
appearance of secondary deformation fields for special classes of materials.
We also find that these latent deformation fields are woken up by normal
stress differences. We present some explicit examples based on the
neo-Hookean, the generalized neo-Hookean, and  the Mooney--Rivlin forms of the
strain-energy density.
Using the simple exact solution found in the neo-Hookean case,
we conjecture that it is advantageous to accompany the usual vertical axial force
by a twisting moment, in order to extrude a cork from the neck of a bottle efficiently.
Then we analyse departures from the neo-Hookean behaviour by exact and by
asymptotic analyses. In that process we are able to give an elegant
and analytic example of secondary (or latent) deformations in the
framework of nonlinear elasticity.

\end{abstract}

\section{Introduction}

Rubbers and elastomers are highly deformable solids which have the
remarkable property of preserving their volume through any deformation. This
permanent isochoricity can be incorporated into the equations of continuum
mechanics through the concept of an \emph{internal constraint}, here the
constraint of \emph{incompressibility}. Mathematically, the formulation of
the constraint of incompressibility has led to the discovery of several
exact solutions in isotropic finite elasticity, most notably to the
controllable or universal solutions of Rivlin and co-workers (see for example
Rivlin (1948)). Subsequently, Ericksen (1954) examined the problem of finding
all such solutions. He found that there are no controllable finite
deformations in isotropic \emph{compressible} elasticity, except for
homogeneous deformations (Ericksen 1955). The impact of that result on the
theory of nonlinear elasticity was quite important and long-lasting, and for
many years a palpable pessimism reigned about the possibility of finding
exact solutions at all for compressible elastic materials. Then Currie \&
Hayes (1981) showed that one could obtain interesting classes of exact
solutions, beyond the homogeneous universal deformations, if one restricted
their attention to certain special classes of compressible materials. A
string of results about the search for exact solutions in nonlinear
elasticity followed. Now a long list exists of classes of exact solutions
which are universal only relative to some special strain-energy functions
(for a recent presentation of such classes see Fu and Ogden (2001).)
These solutions can help us to understand the structure of the theory of
nonlinear elasticity and to complement the celebrated solutions of Rivlin.

In the same vein, some recent efforts focused on determining the maximal
strain energy for which a certain deformation field, fixed a priori, is
admissible. This is a sort of \emph{inverse problem}: find the elastic
materials (that is, the functional form of the strain-energy function) for
which a given deformation field is controllable (that is, for which the
deformation is a solution to the equilibrium equations in the absence of
body forces). A classical example illustrating such an approach is obtained
by considering deformations of \emph{anti-plane shear} type. Knowles (1977)
shows that a non-trivial (non-homogeneous) equilibrium state of anti-plane
shear is not always (universally) admissible, not only for compressible
solids (as expected from Ericksen's result) but also for incompressible
solids (Horgan (1995) gives a survey of anti-plane shear deformations in
nonlinear elasticity). Only for a special class of incompressible materials
(inclusive of the so-called `generalized neo-Hookean materials') is an
anti-plane shear deformation controllable.

Let us consider for example the case of an elastic material filling the
annular region between two coaxial cylinders, with the following boundary
value problem: hold fixed the outer cylinder and pull the inner cylinder by
applying a tension in the axial direction. It is well established that a
solution to this problem, valid for every isotropic incompressible elastic
solid, is obtained by assuming a priori that the deformation field is a pure
axial shear. Now consider the corresponding problem for \emph{non-coaxial}
cylinders, thereby losing the axial symmetry. Then it is clear that we
cannot expect the material to deform as prescribed by a pure axial shear
deformation. Knowles's result tells us that now the boundary value problem
can be solved with a general anti-plane deformation (not axially symmetric)
\emph{only for a subclass} of incompressible isotropic elastic materials. Of
course this restriction does not mean that for a generic material it is not
possible to deform the annular material as prescribed by our boundary
conditions, but rather that in general, these lead to a deformation field
which is more complex than an anti-plane shear. And so, we expect \textit{%
secondary} in-plane deformations.

The theory of non-Newtonian fluid dynamics has generated a
substantial literature about secondary flows, see for example
Fosdick \& Serrin (1973). In solid mechanics it seems that only
Fosdick \& Kao (1978) and Mollica \& Rajagopal (1997) produced
some significant and beautiful examples of secondary deformations
fields for the non-coaxial cylinders problem, although this topic
is clearly of fundamental importance, not only from a theoretical
point of view but also for technical applications.

In this paper we consider a complex deformation field in isotropic
incompressible elasticity, to point out by an explicit example the
situations just evoked and to elaborate on their possible impact on solid
mechanics. Our deformation field takes advantage of the radial symmetry and
therefore we find it convenient to visualize it by considering an elastic
cylinder.

\begin{figure}[tbp]
\centering \mbox{\epsfig{figure=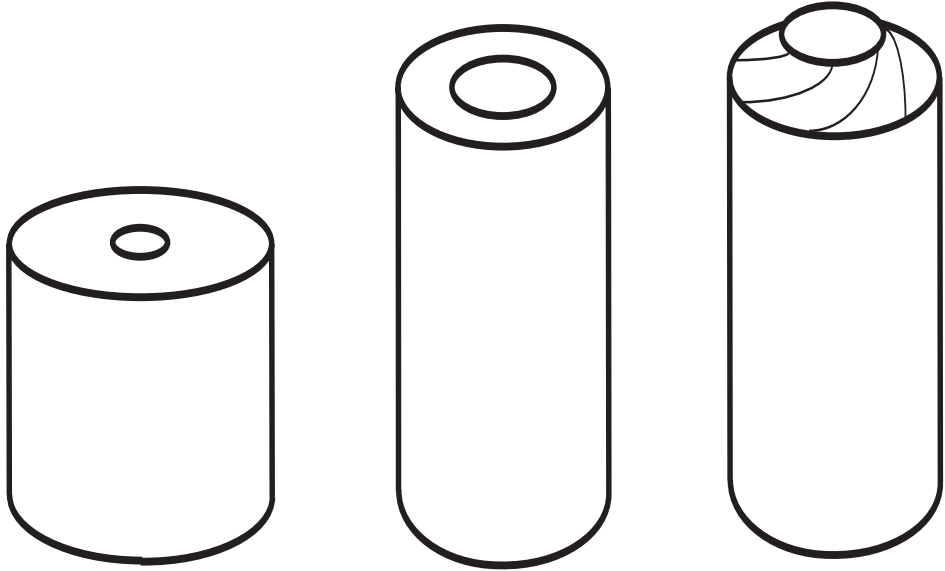,
width=.9\textwidth, height=.55\textwidth}} \caption{Shrink fit of
an elastic tube, followed by the combination of simple torsion and
helical shear. (The figure does not respect scales among the
various deformations)} \label{fig_deformation}
\end{figure}

Let us imagine that a corkscrew has been driven through a cork (the
cylinder) in a bottle.
The inside of the bottleneck is the outer rigid
cylinder and the idealization of the gallery carved out by
the corkscrew constitutes the inner coaxial
rigid cylinder. Our first deformation is purely radial, originated from the
introduction of the cork into the bottleneck and then completed when the
corkscrew penetrates the cork (a so-called \emph{shrink fit problem}, which
is a source of elastic residual stresses here). We call $A$, $B$ the
respective inner and outer radii of the cork in the reference configuration
and $r_1 > A$, $r_2 < B$ their current counterparts. Then we follow with a
\emph{simple torsion} combined to a \emph{helical shear}, in order to model
pulling the cork out of the bottleneck in the presence of a contact force.
Figure \ref{fig_deformation} sketches this deformation.

Of course we are aware of the shortcomings of our modelling with respect to
the description of a `real' cork-pulling problem, because no cork is an
infinitely long cylinder, nor is a corkscrew perfectly straight. In
addition, traditional corks made from bark are anisotropic (honeycomb
mesoscopic structure) and possess the remarkable (and little-known) property
of having an infinitesimal Poisson ratio equal to zero, see the review
article by Gibson \emph{et al.} (1981). However we note that \emph{polymer corks} have
appeared on the world wine market; they are made of elastomers, for which
incompressible, isotropic elasticity seems like a reasonable framework
(indeed the documentation of these synthetic wine stoppers indicates that
they lengthen during the sealing process). We hope that this study provides
a first step toward a nonlinear alternative to the linear elasticity testing
protocols presented in the international standard ISO 9727. We also note
that low-cost \emph{shock absorbers} often consist of a moving metal cylinder,
glued to the inner face of an elastomeric tube, whose outer face is glued to
a fixed metal cylinder (Hill 1975).


The plan of the paper is the following. Section 2 is devoted to
the derivation of the governing equations and to a detailed
description of the boundary value problem. 
In \S3 we specialize the analysis to the \emph{neo-Hookean strain-energy
density}, and find the corresponding exact solution. We use it to
show that it is advantageous to add a twisting moment to an axial
force when extruding a cork from a bottle. The neo-Hookean
strain-energy density is linear with respect to the first
principal invariant of the Cauchy-Green strain tensor. It is much
used in Finite Elasticity theory, although it captures poorly the
basic features of rubber behaviour (Saccomandi 2004). We thus
investigate the consequence of departing from that
strain-energy density. First, in \S4 we consider the
\emph{generalized neo-Hookean strain-energy density}  ---
non-linear with respect to the first principal invariant of the
Cauchy-Green strain tensor --- to show that in this case torsion
is explicitly present in the solution for the axial shear
displacement, but it is a second-order dependence. Next we
consider in \S5 the \emph{Mooney--Rivlin strain-energy
density} --- linear with respect to the first and second principal
invariants of the Cauchy-Green strain tensor --- and find that it is
then also possible to obtain an exact solution to our boundary
value problem. Its expression is too cumbersome to manipulate and
we resort to a small parameter asymptotic expansion from the
neo-Hookean case. Section 6 concludes the paper with some remarks
on the limitations of the semi-inverse method.


\section{Basic equations}


Consider a long hollow cylindrical tube, composed of an isotropic
incompressible nonlinearly elastic material. At rest the tube is in the
region
\begin{equation}
A\leq R\leq B, \qquad 0\leq \Theta \leq 2\pi ,\qquad -\infty \leq Z\leq
\infty,
\end{equation}
where $(R, \Theta ,Z)$ are the cylindrical coordinates associated with the
undeformed configuration, and $A$ and $B$ are the inner and outer radii of
the tube, respectively.


\subsection{Equilibrium equations}


Consider the general deformation obtained by the combination of radial
dilatation, helical shear, and torsion, as
\begin{equation}
r=r(R),\qquad \theta =\Theta +g(R)+\tau Z,\qquad z=\lambda Z+w(R),
\label{deformation}
\end{equation}
where $(r,\theta ,z)$ are the cylindrical coordinates in the deformed
configuration, $\tau $ is the amount of torsion and $\lambda $ is the
stretch ratio in the $Z$ direction. Here, $g$ and $w$ are yet unknown
functions of $R$ only 
(The classical case of pure torsion corresponds to $w = g = 0$, 
see the textbooks by Ogden (1997) or by Atkin and Fox (2007), for instance.)

Hidden inside \eqref{deformation} is the \emph{shrink fit deformation}
\begin{equation}
r=r(R),\qquad \theta =\Theta ,\qquad z=\lambda Z,  \label{deformation1}
\end{equation}
which is \eqref{deformation} without any torsion nor helical shear
($\tau = g = w \equiv 0$).

The physical components of the deformation gradient $\vec{F}$ and of its inverse
$\vec{F}^{-1}$ are then
\begin{equation}
\begin{bmatrix}
r^{\prime } & 0 & 0 \\
rg^{\prime } & r/R & r\tau  \\
w^{\prime } & 0 & \lambda
\end{bmatrix}
,\qquad
\begin{bmatrix}
r\lambda /R & 0 & 0 \\
rw^{\prime }\tau -rg^{\prime }\lambda  & r^{\prime }\lambda  & -rr^{\prime
}\tau  \\
-rw^{\prime }/R & 0 & rr^{\prime }/R
\end{bmatrix}
,  \label{F}
\end{equation}
respectively. Note that we used the incompressibility constraint in order to
compute $\vec{F}^{-1}$; it states that $\text{det }\vec{F}=1$, so that
\begin{equation}
r^{\prime }=\dfrac{R}{\lambda r}.  \label{r'}
\end{equation}

In our first deformation, the cylindrical tube is pressed into a
cylindrical cavity with inner radius $r_{1}>A$ and outer radius $r_{2}<B$.
It follows by integration of the equation above that
\begin{equation}
r(R)=\sqrt{\dfrac{R^2}{\lambda} + \alpha},  \label{r(R)}
\end{equation}
where now
\begin{equation}  \label{alpha_lambda}
\alpha = \frac{B^2 r_1^2 - A^2 r_2^2}{B^2 - A^2}, \qquad \lambda = \frac{B^2
- A^2}{r_2^2 - r_1^2}.
\end{equation}

We compute the physical components of the left Cauchy-Green strain tensor $\vec{B}
\equiv \vec{F F}^t$ from \eqref{F} and find its first three principal invariants $%
I_1 \equiv \text{tr }\vec{B}$, $I_2 \equiv (\text{det }\vec{B})\text{tr }(\vec{B}^{-1})$, and $%
I_3 \equiv \text{det }\vec{B}$ as
\begin{align} \label{I_1_I_2}
& I_1 = (r^{\prime})^2 + (r g^{\prime})^2 + (r/R)^2 + (r \tau)^2 + \lambda^2
+ (w^{\prime})^2,  \notag \\
& I_2 = (r \lambda /R)^2 + (r w^{\prime}\tau - r g^{\prime}\lambda)^2 +
(rw^{\prime}/R)^2 + (R/r)^2 + (1/\lambda)^2 + (R \tau / \lambda)^2,
\end{align}
and of course, $I_3 = 1$.

For a general incompressible hyperelastic solid, the Cauchy stress tensor $\vec{T}$
is related to the strain through
\begin{equation}
\vec{T} = -p \vec{I} + 2 W_1 \vec{B} - 2 W_2 \vec{B}^{-1},
\end{equation}
where $p$ is the Lagrange multiplier introduced by the incompressibility
constraint, $W = W(I_1, I_2)$ is the strain energy density, and $W_i \equiv
\partial W / \partial I_i$. Having computed $\vec{B}^{-1} \equiv (\vec{F}^t)^{-1} \vec{F}^{-1}$
from \eqref{F}, we find that the components of $\vec{T}$ are
\begin{align} \label{components}
& T_{r r} = -p + 2 W_1 (r^{\prime})^2 - 2W_2 \left[ (r \lambda /R)^2 + (r
w^{\prime}\tau - r g^{\prime}\lambda)^2 + (rw^{\prime}/R)^2 \right],  \notag
\\[2pt]
& T_{\theta \theta} = -p + 2 W_1 \left[(rg^{\prime})^2 + (r/R)^2 + (r\tau)^2 %
\right] -2 W_2 (R/r)^2,  \notag \\[2pt]
& T_{z z} = -p + 2 W_1 [\lambda^2 + (w^{\prime})^2] - 2 W_2 \left[%
(1/\lambda)^2 + (R \tau / \lambda)^2 \right],  \notag \\[2pt]
& T_{r \theta} = 2 W_1 (r r^{\prime}g^{\prime}) - 2 W_2 (w^{\prime}\tau -
g^{\prime}\lambda)R,  \notag \\[2pt]
& T_{r z} = 2 W_1 (r^{\prime}w^{\prime}) -2 W_2 \left[r R g^{\prime}\tau - r
R w^{\prime}\tau^2/\lambda - r w^{\prime}/(\lambda R) \right],  \notag \\%
[2pt]
& T_{\theta z} = 2 W_1 (r w^{\prime}g^{\prime}+ r \lambda \tau) + 2 W_2
(r^{\prime}R \tau).
\end{align}

Finally the equilibrium equations, in the absence of body forces, are: $%
\text{div } \vec{T}= \mathbf{0}$; for fields depending only on the radial
coordinate as here, they reduce to
\begin{equation}
\dfrac{\text{d} T_{r r}}{\text{d} r} + \dfrac{T_{r r} - T_{\theta \theta}}{r}
= 0, \qquad \dfrac{\text{d} T_{r \theta}}{\text{d} r} + \frac{2}{r} T_{r
\theta} = 0, \qquad \dfrac{\text{d} T_{r z}}{\text{d} r} + \frac{1}{r} T_{r
z}=0.  \label{equaRidotte}
\end{equation}

\subsection{Boundary conditions}

Now consider the inner face of the tube: we assume that it is subject to a
vertical pull,
\begin{equation} \label{BC_A}
T_{r z}(A) = T_0^A, \qquad  T_{r \theta}(A) = 0,
\end{equation}
say. Then we can integrate the second and third equations of equilibrium %
\eqref{equaRidotte}$_{2,3}$; we find that
\begin{equation}
T_{r z}(r) = \frac{r_1}{r}T_0^A, \qquad T_{r\theta}(r)=0.
\label{Trtheta,Trz}
\end{equation}

The outer face of the tube (in contact with the glass in the cork/bottle
problem) remains fixed, so that
\begin{equation}
w(B)=0,\qquad g(B)=0,\qquad T_{rr}(B)=T_{0},  \label{condizioni al bordo}
\end{equation}
say.
In addition to the axial traction applied on its inner face, the tube is
subject to a resultant axial force $N$ (say) and a resultant moment $M$
(say),
\begin{equation}
N = \int_0^{2\pi } \int_{r_1}^{r_2} T_{zz} r \text{d}r \text{d}\theta,
\qquad M = \int_0^{2\pi } \int_{r_1}^{r_2} T_{\theta z} r^2 \text{d}r \text{d}\theta.
\label{forza assiale e momento}
\end{equation}
Note that the traction $T_{0}$ of \eqref{condizioni al bordo}
is not arbitrary but is instead determined by the
\emph{shrink fit pre-deformation} \eqref{deformation1},
by requiring that $N=0$ when $T_0^A = \tau = g = w \equiv 0$
(This process is detailed in the next section for the neo-Hookean material.)
 Therefore $T_{0}$ is
connected with the stress field experienced by the cork when it is
introduced in the bottleneck.

In the rest of the paper we aim at presenting results in dimensionless
form. To this end we normalize the strain-energy density $W$ and the Cauchy
stress tensor $\vec{T}$ with respect to $\mu $, the infinitesimal shear modulus;
hence we introduce $\overline{W}$ and $\overline{\vec{T}}$ defined by
\begin{equation}
\overline{W}=\dfrac{W}{\mu },\qquad \overline{\vec{T}}=\dfrac{\vec{T}}{\mu }.
\end{equation}
Similarly we introduce the following non-dimensional variables,
\begin{equation}
\eta =\dfrac{A}{B},\quad \overline{R}=\dfrac{R}{B},\quad \overline{r}_{i}=%
\dfrac{r_{i}}{B},\quad \overline{w}=\dfrac{w}{B},\quad \overline{\alpha }=%
\dfrac{\alpha }{B^{2}},\quad \overline{\tau }=B\tau ,
\end{equation}
so that $\eta \leq \overline{R}\leq 1$. Also, we find from \eqref{alpha_lambda} that
\begin{equation} \label{alpha_lambda_1}
\overline{\alpha }=\dfrac{\overline{r}_{1}^{2}-\eta ^{2}\overline{r}_{2}^{2}%
}{1-\eta ^{2}},\qquad \lambda =\dfrac{1-\eta ^{2}}{\overline{r}_{2}^{2}-%
\overline{r}_{1}^{2}}.
\end{equation}

Turning to our cork or shock absorber problems, we imagine that the inner
metal cylinder is introduced into a pre-existing cylindrical cavity (this
precaution ensures a one-to-one correspondence of the material points
between the reference and the current configurations). In our upcoming
numerical simulations, we take $A= B/10$ so that $\eta = 0.1$; we consider
that the outer radius is shrunk by 10\%: $r_2 = 0.9 B$, and that the inner
radius has doubled: $r_1 = 2 A$; finally
we apply a traction whose magnitude is half the infinitesimal shear modulus:
$|T_0^A| = \mu / 2$. This gives
\begin{equation}  \label{numerics}
\overline{\alpha } \simeq 3.22 \times 10^{-2}, \qquad \lambda \simeq 1.286,
\qquad \overline{T}_0^A = -0.5.
\end{equation}

%
%
%

At this point it is possible to state clearly our main
observation. A first glance at the boundary conditions, in
particular at the requirements
that $g$ be zero on the outer face of the tube, gives the expectation that $%
g \equiv 0$ everywhere is a solution to our boundary value
problem, at least for some simple forms of the constitutive
equations. In what follows we find that for the so-called
`neo-Hookean' solids, $g \equiv 0$ is indeed a solution, whether
there is a torsion $\tau$ or not. However if the solid is not
neo-Hookean, then it is necessary that $g \ne 0$ when  $\tau \ne
0$, and the picture becomes more complex. For this reason, we
classify as `purely academic' the question: \emph{which is the most
general strain-energy density for which it is possible to solve
the above boundary value problem with $g \equiv 0$} ?
Indeed there is no `real world' material whose behaviour is 
ever going to be described \emph{exactly} by that strain-energy density
(supposing it exists).
Instead a more pertinent issue to raise for `real word
applications' is whether we are able to evaluate the importance of latent 
(secondary) stress fields, because they are bound to be woken up (triggered)
by the deformation.

\section{neo-Hookean materials}

First we consider the special strain energy density which generates the
class of neo-Hookean materials, namely
\begin{equation}  \label{neo}
W = (I_1 - 3)/2, \qquad \text{so that} \qquad 2 W_{1} = 1, \qquad W_2=0.
\end{equation}
Note that here and hereafter, we use the non-dimensional quantities
introduced previously, from which we drop the overbar for convenience. Hence
the components of the (non-dimensional) stress field in a neo-Hookean
material reduce to
\begin{align}
& T_{rr} = -p + (r^{\prime})^2, & & T_{\theta \theta} = -p + (r
g^{\prime})^2 + (r/R)^2 + (r \tau)^2,  \notag \\
& T_{z z} = -p + \lambda^2 + (w^{\prime})^2, & & T_{r \theta} = r
r^{\prime}g^{\prime},  \notag \\
& T_{r z} = r^{\prime}w^{\prime}, & & T_{\theta z} = r g^{\prime}w^{\prime}+
r \lambda \tau.  \label{componentiTconW_2=0}
\end{align}
Substituting into \eqref{Trtheta,Trz} we find that
\begin{equation}
w^{\prime}= \lambda r_1 T_0^A/R, \qquad g^{\prime}= 0,
\label{sistemag'w'dimensionless}
\end{equation}
and by integration, using \eqref{condizioni al bordo}, that
\begin{equation}
w = \lambda r_1 T_0^A \ln R, \qquad g=0.  \label{w,gnel_neo-Hookeano}
\end{equation}

In Figure 2a, we present a rectangle in the
tube at rest. It is delimited by $0.1 \le R \le 1.0$ and $0.0 \le Z \le 1.0$. 
Then it is subject to the deformation corresponding to the numerical
values of \eqref{numerics}. 
To generate Figure 2b, we computed
the resulting shape for a neo-Hookean tube, using \eqref{deformation}, %
\eqref{r(R)}, and \eqref{w,gnel_neo-Hookeano}.
\begin{figure}[tbp]
\centering
\mbox{\subfigure{\epsfig{figure=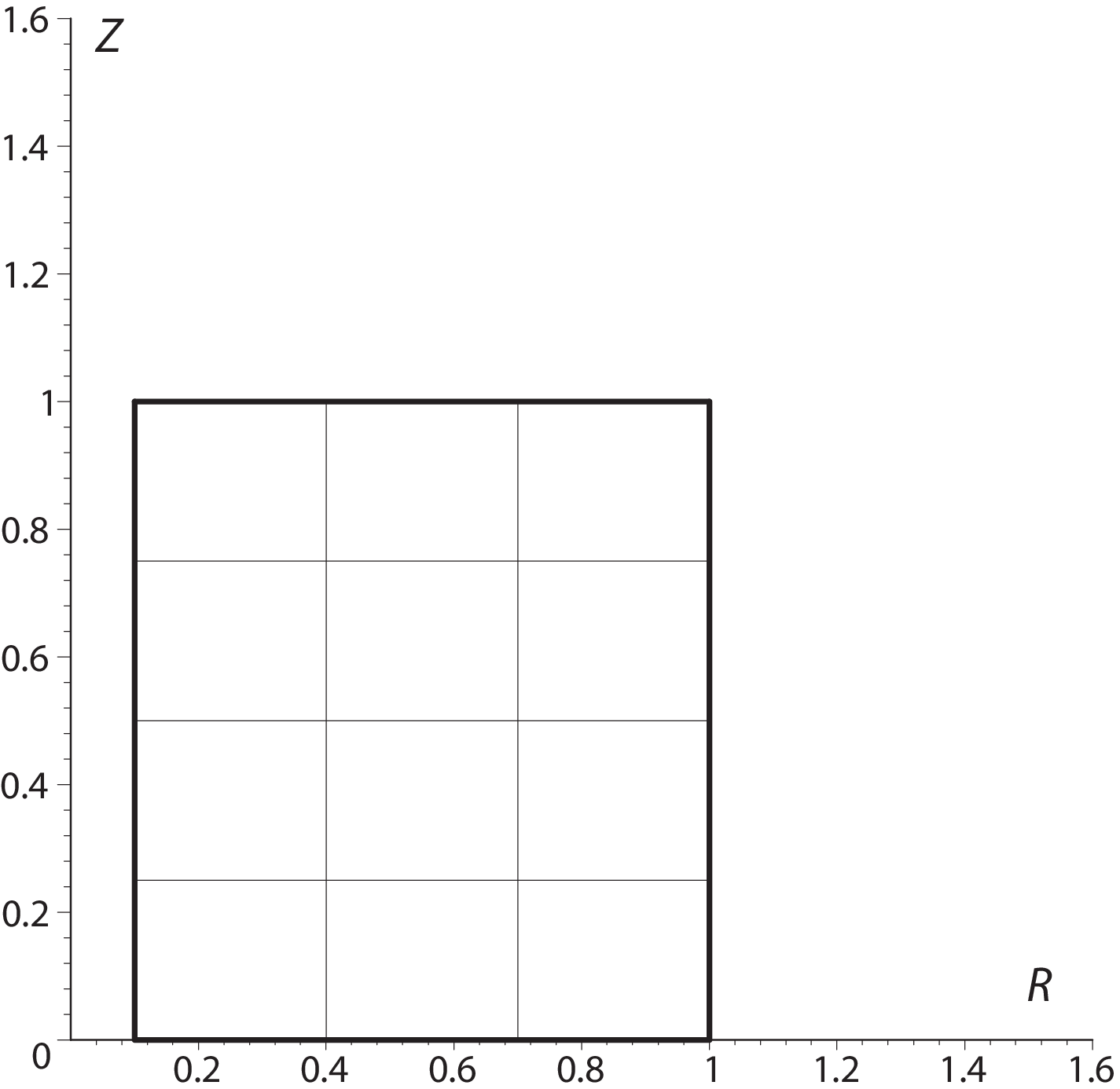, width=.45\textwidth, height=.45\textwidth}}
  \quad
     \subfigure{\epsfig{figure=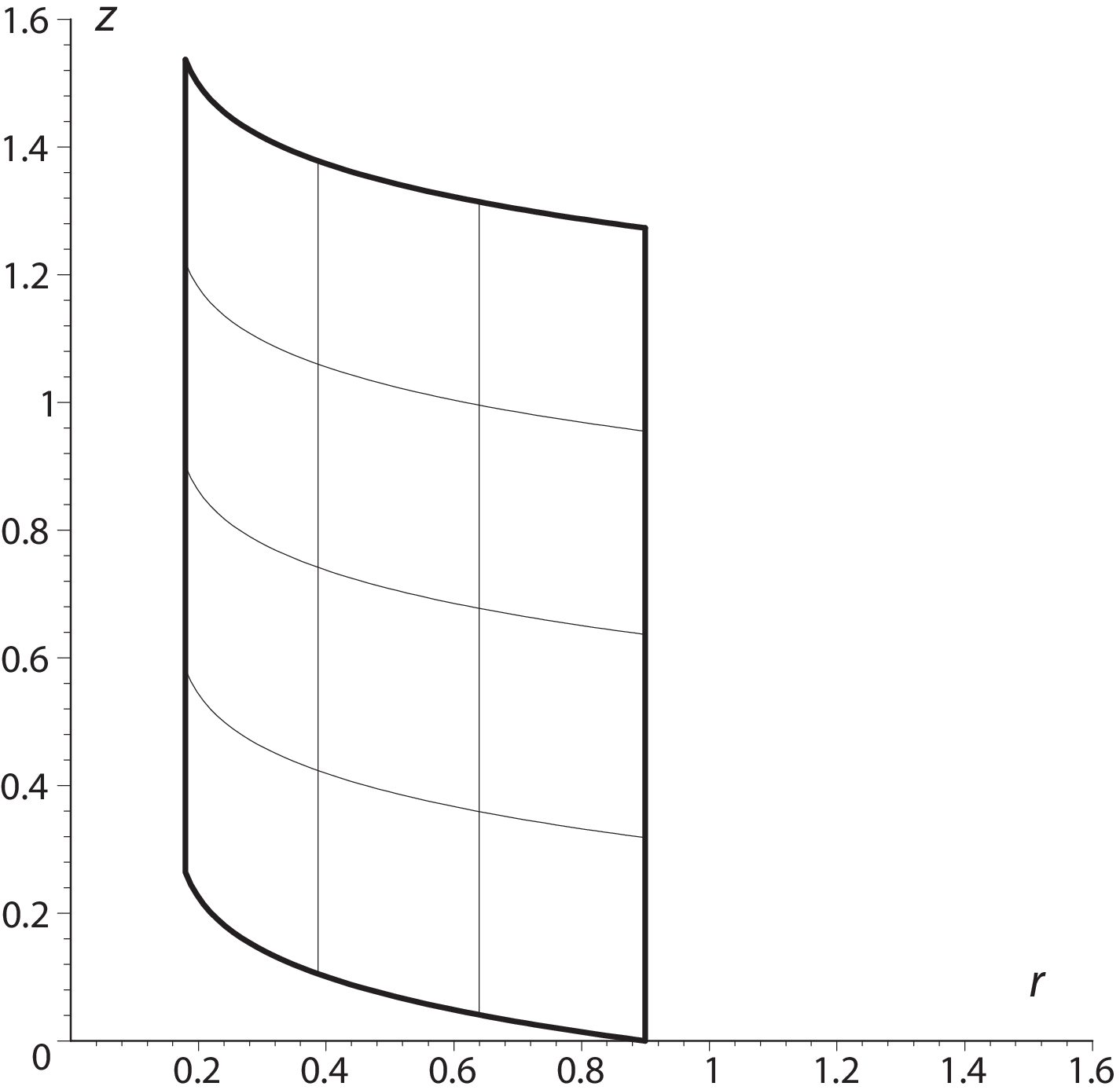, width=.45\textwidth, height=.45\textwidth}}}
\caption{Pulling on the inside face of a neo-Hookean tube. Here the vertical
axis is the symmetry axis of the tube. }
\label{figure_neo}
\end{figure}

Now that we know the ful deformation field, see \eqref{deformation} and
\eqref{w,gnel_neo-Hookeano}, we can compute $T_{rr} - T_{\theta \theta }$
from \eqref{componentiTconW_2=0} and deduce $T_{rr}$ by
integration of \eqref{equaRidotte}$_{1}$, with initial condition
\eqref{condizioni al bordo}$_{3}$.
Then the other field quantities follow from the rest of \eqref{componentiTconW_2=0}.
In the end we find in turn that
\begin{align}
& T_{rr} =
  \frac{1}{2 \lambda}\left\{
     \ln \frac{\lambda r_2^2 R^2}{R^2 + \alpha \lambda}
       + (R^2 - 1) \left[\frac{\alpha}{r_2^2 \left(R^2 + \alpha \lambda \right)} + \tau^2 \right]
                        \right\}
+  T_{0},
\notag \\[2pt]
& T_{\theta \theta}  = T_{rr}
    + \left(\frac{R^2}{\lambda} + \alpha \right)
        \left(\frac{1}{R^2} + \tau^2\right)
          - \frac{R^2}{\lambda (R^2 + \alpha \lambda)},
\notag \\[2pt]
& T_{zz}  =  T_{rr}
       + \lambda^2 \left(1+ \frac{r_1^2 (T_0^A)^2}{R^2} \right)
         - \frac{R^2}{\lambda(R^2 + \alpha \lambda)},
\label{ttt}
\end{align}
(where we used the identity $1 + \alpha \lambda = \lambda r_2^2$,
see \eqref{r(R)} with $R=1$), and that
\begin{equation} \label{ttz}
T_{r\theta }=0,
\qquad
T_{rz}=\dfrac{r_{1}}{\sqrt{\dfrac{R^{2}}{\lambda} +\alpha }}T_{0}^{A},
\qquad
T_{\theta z}=\lambda \tau \sqrt{\dfrac{R^{2}}{\lambda }+\alpha }.
\end{equation}
The constant $T_0$ is fixed by the shrink fit pre-deformation \eqref{deformation1},
imposing that $N=0$ when $\tau = g = w = T_0^A \equiv 0$, or
\begin{equation}
(T_0 + \lambda^2)(1 - \eta^2) +
  \frac{1}{\lambda}\int_{\eta}^{1}\left[
     \ln \frac{\lambda r_2^2 R^2}{R^2 + \alpha \lambda}
         + \frac{\alpha (R^2 - 1)}{r_2^2(R^2 + \alpha \lambda)}
       - \frac{2 R^2}{R^2 + \alpha \lambda}
            \right] R \text{d}R = 0.
\end{equation}
Using this, and \eqref{forza assiale e momento}, \eqref{ttt}, \eqref{ttz},
we find the following expressions for the resultant moment,
\begin{equation}
M=\pi (r_{2}^{4}-r_{1}^{4})\lambda \tau /2,  \label{moment}
\end{equation}
and for the axial force,
\begin{equation}
N =  2 \pi \lambda r_1^2 |\ln \eta| \left(T_0^A \right)^2 -
 \frac{\pi}{4}(r_2^2 - r_1^2)^2 \tau^2.
   \label{nn}
\end{equation}

We now have a clear picture of the response of a neo-Hookean solid to the
deformation \eqref{deformation}, with the boundary conditions of \S2$\,b$. 
First we saw that here the contribution $g(R)$
is not required for the azimuthal displacement,
whether there is a torsion $\tau$ or not. Also, if a moment $M\neq 0$
is applied, then the tube suffers an amount of torsion $\tau \neq 0$
proportional to $M$. On the other hand, if the tube is pulled by the
application of an axial force only ($N\neq 0$) and no moment ($M=0$), then $%
\tau =0$ and no azimuthal shear occurs at all.

When we try to apply our results to the extrusion of a cork from
the neck of a bottle, the following remarks seem to be relevant.
From the elementary theory of Coulomb friction, it is known
that the pulled cork starts to move when, in modulus, the friction
force exerted on the neck surface is equal to the normal force
times the coefficient of static friction. In our case this means
that 
\begin{equation}
\sqrt{\left| T_{rz}(1)\right|^2 + \left| T_{r \theta}(1)\right|^2} =
  f_{S}\left| T_{rr}(1)\right| = f_{S}\left| T_0 \right|,  \label{cc}
\end{equation}
where $f_{S}$ is the coefficient of static friction.
Using \eqref{BC_A} and \eqref{Trtheta,Trz},
we find that the elements
 of the left handside of this inequality are
\begin{equation}
 T_{rz}(1) = ( r_1 /r_2) T_0^A, \qquad
  T_{r \theta}(1) =  0.  \label{cc2}
\end{equation}
Now, our main concern is to understand if it is better to
apply a moment $M \ne 0$ when uncorking a bottle, than to pull only.
To address this question we note that the left handside of inequality \eqref{cc}
increases when $|T_0^A|$ increases;
on the other hand, combining \eqref{moment} and \eqref{nn}, we have
\begin{equation}
 (T_0^A)^2 =
   \left[N + \frac{1}{\pi \lambda^2 (r_1^2 + r_2^2)^2} M^2 \right]/(2 \pi \lambda r_1^2 |\ln \eta|),
\label{cc2bis}
\end{equation}
It is now clear, that for a fixed value of $T_0^A$, in the case $M
\neq 0$, it is necessary to apply an axial force whose intensity
is less than the one in the case $M=0$. Moreover, the equation
above shows that $\left(T_0^A\right)^2$ grows linearly with $N$
but quadratically with $M$. With respect to efficient
cork-pulling, the conclusion is that adding a twisting moment to a
given pure axial force is more advantageous than solely increasing
the vertical pull. Moreover, we observe that a moment is applied
by using a lever and this is always more convenient from an
energetic point of view.

Recall that we made several simplifying assumptions to reach these results:
not only infinite axial length, incompressibility, and
isotropy, but also the choice of a truly special strain energy density.
In the next two sections we depart from the neo-Hookean model.

\section{Generalized neo-Hookean materials}

As a first broadening of the
neo-Hookean strain-energy density \eqref{neo}, we consider
\emph{generalized neo-Hookean materials}, for which the
strain-energy density is a nonlinear function of the first invariant $I_1$ only,
\begin{equation} \label{W(I1)}
 W = \widehat{W}(I_1),
\end{equation}
say.
To gain access to the Cauchy stress components in this context,
it suffices to take $W_2 = 0$ and $W_1 = \widehat{W}'$ in equations \eqref{components}.
In particular,
$T_{r \theta} = 2 r r' g'  \widehat{W}'$,
and the integrated equation of equilibrium \eqref{Trtheta,Trz}$_2$ gives $g'=0$.
Integrating, with \eqref{BC_A} as an initial value, we find that
\begin{equation}
 g \equiv 0.
\end{equation}
Hence, just as in the neo-Hookean case,
azimuthal shear can be avoided altogether, whether there is a torsion $\tau$ or not.
We are left with an equation for the axial shear, namely \eqref{Trtheta,Trz}$_1$,
which here reads
\begin{equation} \label{generalized}
 2  \widehat{W}'(I_1) w'(R) = \frac{\lambda r_1}{R} T_0^A.
\end{equation}
Obviously the same steps as those taken for neo-Hookean solids may be followed here
for any given strain energy density \eqref{W(I1)}, but
now by resorting to a numerical treatment.
Horgan \& Saccomandi (2003$a$) show, through some specific examples of hardening generalized
neo-Hookean solids, how rapidly involved the analysis becomes,
even when there is only helical shear and no shrink fit.

Instead we simply point out some striking differences between our
present situation and the neo-Hookean case.
We remark that $I_1$ is of the form \eqref{I_1_I_2}$_1$ at $g \equiv 0$
that is,
\begin{equation}
I_1 = \lambda^2 + \frac{R^2}{\lambda (R^2 + \alpha \lambda)}
   + \left(\frac{R^2}{\lambda} + \alpha \right)
       \left(\frac{1}{R^2} + \tau^2 \right)  +  [w'(R)]^2.
\end{equation}
It follows that \eqref{generalized} is a \emph{nonlinear}
differential equation for $w'$,
in contrast with the neo-Hookean case.
Another contrast is that the axial shear $w$ is now intimately
coupled to the torsion
parameter $\tau$, and that this dependence is a \emph{second-order effect}
($\tau$ appears above as $\tau^2$).

A similar problem where the azimuthal shear has not been ignored,
but the axial shear has been considered null i.e. $w \equiv 0$ has
been recently considered by Wineman (2005).

\section{Mooney--Rivlin materials}

In this section we specialize the general equations of \S2 to the
Mooney--Rivlin form of the strain-energy density, which in its
non-dimensional form reads
\begin{equation}  \label{mooney}
W = \frac{I_1 -3 + m (I_2 - 3)}{2(1+m)}, \quad \text{so that} \quad 2 W_1 =
\frac{1}{1 + m} \quad 2 W_2 = \frac{m}{1 + m},
\end{equation}
where $m > 0$ is a material parameter, distinguishing the Mooney--Rivlin
material from the neo-Hookean material \eqref{neo},
and also allowing a dependence on the second principal strain invariant $I_2$,
in contrast to the generalized neo-Hookean solids of the previous section.

Then the integrated equations of equilibrium \eqref{Trtheta,Trz} read
\begin{align}
& \left(R + m \tau^2 r^2 R + m r^2/R \right) w^{\prime}- (m \tau \lambda r^2
R) g^{\prime}= (1 + m) \lambda r_1 T_0^A,  \notag \\
& (m \tau \lambda)w^{\prime}- (1 + m \lambda^2) g^{\prime}= 0.
\label{sistemag'w'}
\end{align}

First we ask ourselves if it is possible to avoid torsion during the pulling
of the inner face. Taking $\tau =0$ above gives
\begin{equation}
(R + m r^2/R) w^{\prime}= (1 + m) \lambda r_1 T_0^A, \qquad g^{\prime}= 0.
\end{equation}
It follows that here it is indeed possible to solve our boundary value problem. We
find
\begin{equation}
w = \lambda r_1 T_0^A \ \dfrac{\lambda (1+m)}{2 (\lambda +m)} \ln \left[
\frac{ m \alpha \lambda + (\lambda + m) R^2} {m \alpha \lambda + (\lambda +
m)}\right], \qquad g = 0.
\end{equation}

However if $\tau \neq 0$, then it is necessary that $g \ne 0$, otherwise %
\eqref{sistemag'w'}$_2$ gives $w^{\prime}=0$ while \eqref{sistemag'w'}$_1$
gives $w^{\prime}\ne 0$, a contradiction.
This constitutes the first departure from the neo-Hookean and generalized neo-Hookean
behaviours:
\emph{torsion ($\tau \ne 0$) is necessarily accompanied by azimuthal shear ($g \ne 0$)}.

In the case $\tau \neq 0$, we
introduce the function $\Lambda = \Lambda(R)$ defined as
\begin{equation}
\Lambda(R) = (R + m r^2/R)(1 + m \lambda^2) + m \tau^2 r^2 R,
\end{equation}
(recall that $r = r(R)$ is given explicitly in \eqref{r(R)}.) We then solve
the system \eqref{sistemag'w'} for $w^{\prime}$ and $g^{\prime}$ as
\begin{equation}
w^{\prime}= (1 + m)(1 + m \lambda^2) \lambda \dfrac{T_0^A}{\Lambda(R)} r_1,
\qquad g^{\prime}= m (1 + m) \lambda^2 \dfrac{T_0^A}{\Lambda(R)} \tau r_1,
\label{pippo}
\end{equation}
making clear the link between $g$ and $\tau$.
Thus for the Mooney--Rivlin material, the azimuthal shear $g$
is a \textit{latent} mode of deformation;
it is \emph{woken up} by any amount of torsion $\tau$.
Recall that at first sight, the azimuthal shear component
of the deformation \eqref{deformation} seemed inessential to satisfy the boundary conditions,
especially in view of the boundary condition $g(1) = 0$.
However, a non-zero $W_2$ term in the constitutive equation clearly
couples the effects of a torsion and of an azimuthal shear, as displayed
explicitly by the presence of $\tau$ in the expression for $g'$ above.

It is perfectly possible to integrate equations \eqref{pippo} in
the general case, but to save space we do not reproduce the
resulting long expressions. 
With them, we generated the deformation field picture of Figure 3a and Figure 3b. 
There we took the numerical values of \eqref{numerics} for $\alpha$, $\lambda$,
$T_0^A$; we took a Mooney--Rivlin solid with $m = 5.0$; we imposed
a torsion of amount $\tau = 0.5$; and we looked at the deformation
field in the plane $Z = 1$ (reference configuration) and $z =
\lambda$ (current configuration).
\begin{figure}[tbp]
\centering
\mbox{\subfigure{\epsfig{figure=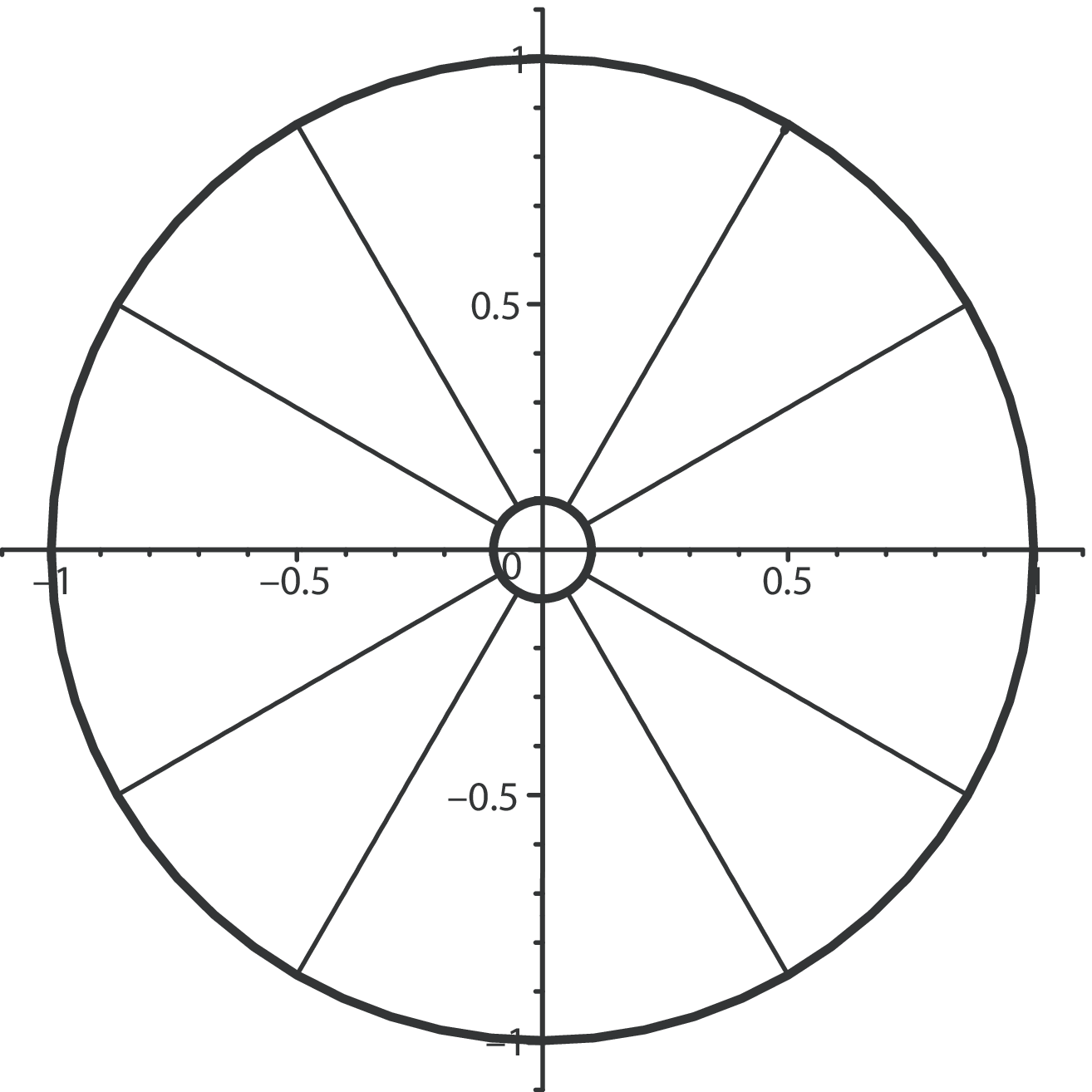, width=.45\textwidth, height=.45\textwidth}}
  \quad
     \subfigure{\epsfig{figure=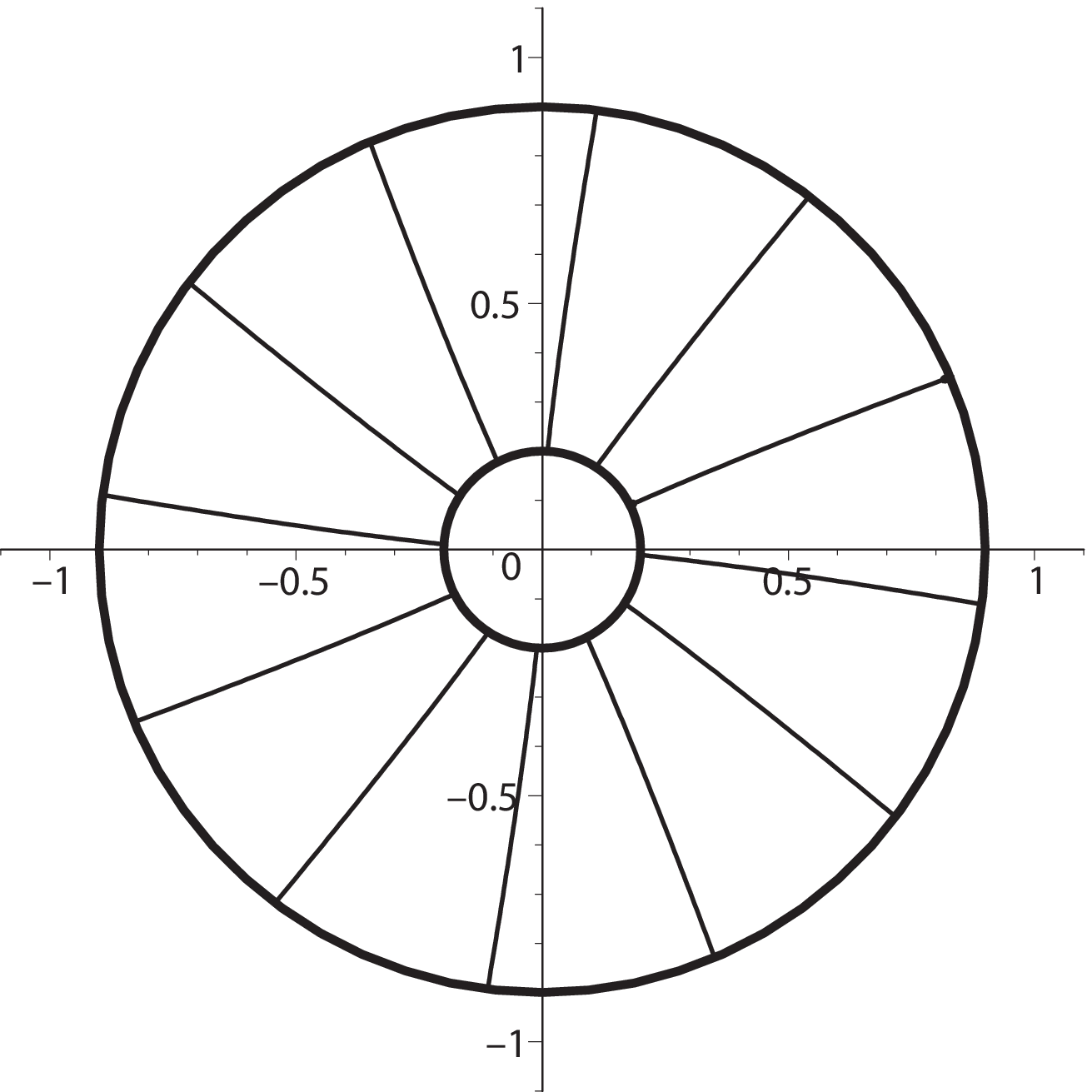, width=.45\textwidth, height=.45\textwidth}}}
\caption{Pulling on the inside face of a Mooney--Rivlin tube, with
a clockwise torsion.} \label{mooneyb}
\end{figure}

Although the secondary fields appear to be slight in the picture, they are nonetheless
truly present and cannot be neglected.
To show this we consider a perturbation method to obtain
simpler solutions and to understand the effect of the coupling, by taking $m$ small.
Then integrating \eqref{pippo}, we find at first order that
\begin{align}\label{pippo2}
& \dfrac{w}{r_{1}T_{0}^{A}} \simeq  (1+m)\lambda \ln {R}
  -\dfrac{1}{2}m \left[ \tau^2 R^2 + 2 (1 + \tau^2 \alpha \lambda) \ln R
      -\alpha \lambda/R^2-\tau ^{2}+\alpha \lambda \right], \notag \\
& \dfrac{g}{r_1 T_0^A} \simeq \lambda^2 \tau m \ln {R}.
\end{align}
Hence, the secondary field $g$ exists even for a nearly neo-Hookean solid
(if $m$ is small, then $g$ of order $m$.)
Interestingly we also note that the azimuthal shear $g$ in \eqref{pippo2} varies
in a homogeneous
and linear manner with respect to the torsion parameter $\tau$ and in a quadratic manner
with respect to the axial stretch $\lambda$, showing that that the presence of this secondary
deformation field cannot be neglected when the effects of the
prestress and of the torsion are both taken into account.
To complete the picture, we use the first-order approximations
\begin{equation}
2W_{1} \simeq 1-m, \qquad 2W_{2} \simeq m,
\end{equation}
to obtain the stress field as
\begin{align}
& T_{rr} \simeq
    -p + \left(1-m\right) (r^{\prime })^{2}
     - m\left\{ (r\lambda
/R)^{2}+\left[ (r\tau )^{2}+(r/R)^{2}\right] \left( \lambda
r_{1}T_{0}^{A}\right) ^{2}/R^{2}\right\},
\notag \\[2pt]
& T_{\theta \theta }\simeq -p+\left( 1-m\right) \left[ (r/R)^{2}+(r\tau )^{2}%
\right] -m(R/r)^{2},
 \notag \\[2pt]
& T_{zz}\simeq -p+\left(\lambda T_0^A r_1\right)^2
 \left[\left(1+2m\lambda^2\right)\frac{1}{R^2}
  - \frac{2}{R}\left(\frac{\tau^2r^2}{R}
   + \frac{r^2}{R^3}+\frac{\lambda^2}{R}-\frac{1}{2R}\right)m
    \right]
\notag \\
& \hspace{2.02cm}+\left( 1-m\right) \lambda ^{2} -m\left[ (1/\lambda
)^{2}+(R\tau /\lambda )^{2}\right] ,  \notag \\[2pt]
& T_{r\theta }\simeq rr^{\prime }g^{\prime }-m\lambda r_{1}T_{0}^{A}\tau ,
\notag \\[2pt]
& T_{rz}\simeq \left( 1-m\right) (r^{\prime }w^{\prime })+m\lambda
r_{1}T_{0}^{A}\left[ r R\tau ^{2}/\lambda +r/(\lambda R)\right] /R,  \notag \\%
[2pt]
& T_{\theta z}\simeq \left( 1-m\right) r\lambda \tau +\lambda
rr_{1}T_{0}^{A}g^{\prime }/R+m(r^{\prime }R\tau ).
\end{align}

Using this stress field it is straightforward, but long and cumbersome,
to derive the analogue for a Mooney--Rivlin solid with a small $m$ of 
relation \eqref{cc2bis} (which was established for neo-Hookean solids.) 
However nothing truly new is gained from these complex
formulas with respect to the simple neo-Hookean case, and 
we do not pursue this aspect any further.


\section{Conclusion}

In non-Newtonian fluid mechanics and in turbulence theory, the existence of
shear-induced normal stresses on planes transverse to the direction of shear
is at the root of some important phenomena occurring in the flow of fluid
down pipes of non-circular cross section (Fosdick \& Serrin 1973). In
other words, pure parallel flows in tubes without axial symmetry are
possible only when we consider the classical theory of Navier-Stokes
equations or the linear theory of turbulence or tubes of circular cross
section.

In nonlinear elasticity theory, similar phenomena are reported. Hence
Fosdick \& Kao (1978) and Mollica  \&  Rajagopal (1997) show that for
general isotropic incompressible materials, an anti-plane shear deformation
of a cylinder with non axial-symmetric cross section causes a secondary
in-plane deformation field, because of normal stress differences. Horgan \&
Saccomandi (2003$b$) give a detailed discussion of how the anti-plane shear
deformation field couples with the in-plane deformation field in a
generalized neo-Hookean solid.

The appearance of what we called here \emph{latent deformations} is quite
general and common. For example it is known in compressible nonlinear
elasticity that pure torsion is possible only in a special class of
materials, but we know that torsion plus a radial displacement is possible
in all compressible isotropic elastic materials (Polignone \& Horgan 1991) 
(Here we signal that `possible in all materials' is not equivalent
to `universal', because the corresponding radial deformation differs from
one material to another.)

In this paper we give an example where axial symmetry holds, where the
boundary conditions suggest that an axial shear deformation field is
sufficient to solve the boundary value problem, and where nevertheless, the
normal stress difference wakes up a latent azimuthal shear deformation.
Moreover, because we are able to find some explicit exact solutions by some
perturbation techniques, we are able to evaluate the \textit{importance} of
the latent deformation. Indeed, we show that if a certain constitutive
parameter $m$ (distinguishing a neo-Hookean solid from a Mooney--Rivlin
solid) is zero or if the torsion parameter $\tau$ is zero, then the solution
to the boundary value problem can be found only in terms of the axial shear
deformation field; if these two parameters are not zero --- even if they are
small --- then the latent mode of deformation is quantitatively appreciable.

In conclusion we suggest that it is not crucial to determine the class of
materials for which a given deformation field is possible. Rather it is
crucial to classify all the latent deformations associated with a given
deformation field in such a way that this field is controllable for the
entire class of materials. Indeed, no \emph{real} material, even when we
accept that its mechanical behaviour is purely elastic, is ever going to be
described exactly by a special choice of strain-energy. Looking for special
classes of materials for which special deformations fields are admissible
may mislead us in our understanding of the nonlinear mechanical behaviour of
materials.

\begin{figure}[ht]
\centering
\mbox{\epsfig{figure=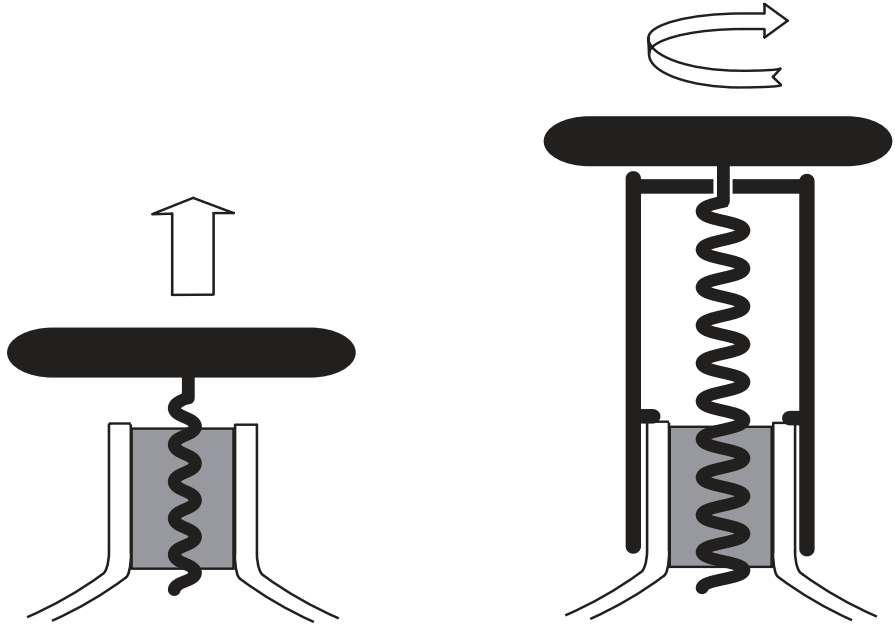, width=.85\textwidth,
height=.6\textwidth}}
\caption{There are two main types of corkscrews: one that relies on pulling only (left)
and one that adds a twist to the cork-pulling action (right).
The analysis developed in this paper indicates that the second type is more efficient.}
\label{fig_corkscrews}
\end{figure}

To finish the paper on a light note, we evoke a classic wine party dilemma:
which kind of corkscrew system requires the least effort to uncork a bottle?
Figure 4 sketches the two working principles commonly found
in commercial corkscrews.
The most common (on the left) relies on pulling only (directly or through levers)
and the other type (on the right) relies on a combination of pulling and twisting.
Notwithstanding the shortcomings of this paper's modelling with respect to an actual
uncorking,
the authors are confident that they have provided a scientific argument to those
wine amateurs who favour the second type of corkscrews over the first.


\end{document}